\documentstyle[prd,aps,floats,graphicx]{revtex}

\begin{document}
\preprint{astro-ph/0304446}
\draft

%
%
\input epsf
\renewcommand{\topfraction}{0.8}
\twocolumn[\hsize\textwidth\columnwidth\hsize\csname
@twocolumnfalse\endcsname

\title{Effect of realistic astrophysical inputs on the phase and
shape of the WIMP annual modulation signal} 
\author{Anne M.~Green}
\address{Physics Department, Stockholm University, Stockholm, S106 91,
Sweden} 
\date{\today} 
\maketitle
\begin{abstract}
The orbit of the Earth about the Sun produces an annual modulation in
the WIMP direct detection rate. If the local WIMP velocity
distribution is isotropic then the modulation is roughly sinusoidal
with maximum in June, however if the velocity distribution is
anisotropic the phase and shape of the signal can change. Motivated by
conflicting claims about the effect of uncertainties in the local
velocity distribution on the interpretation of the DAMA annual
modulation signal (and the possibility that the form of the modulation
could be used to probe the structure of the Milky Way halo), we study
the dependence of the annual modulation on various astrophysical
inputs.  We first examine the approximations used for the Earth's
motion about the Sun and the Sun's velocity with respect to the
Galactic rest frame. We find that overly simplistic assumptions lead
to errors of up to ten days in the phase and up to tens of per-cent in
the shape of the signal, even if the velocity distribution is
isotropic. Crucially, if the components of the Earth's velocity
perpendicular to the motion of the Sun are neglected, then the change
in the phase which occurs for anisotropic velocity distributions is
missed.  We then examine how the annual modulation signal varies for
physically and observationally well-motivated velocity
distributions. We find that the phase of the signal changes by up to
20 days and the mean value and amplitude change by up to tens of
per-cent.
\end{abstract}


\vskip2pc]

\section{Introduction}

Arguably the best motivated non-baryonic dark matter candidate is the
neutralino (the lightest supersymmetric particle), and direct
detection experiments are just reaching the sensitivity required to
probe the relevant region of parameter space~\cite{lars}. Since the
expected event rates are so small ( ${\cal O} (10^{-5} - 10)$ counts
${\rm kg^{-1} day^{-1}}$ see e.g. Refs.~\cite{jkg,ls}) distinguishing
a putative Weakly Interacting Massive Particle (WIMP) signal from
backgrounds due to, for instance, neutrons from cosmic-ray induced
muons or natural radioactivity, is crucial.  The Earth's motion about
the Sun provides two potential WIMP smoking guns: i) an annual
modulation~\cite{amtheory} and ii) a strong direction
dependence~\cite{dirndep} of the event rate. In principle the
dependence of the differential event rate on the atomic mass of the
detector (see e.g. Refs.~\cite{jkg,ls}) is a third possibility,
however this would require good control of systematics for detectors
composed of different materials. While direction sensitive detectors
probably offer the best long-term prospects for the unambiguous
detection of WIMPs, and development of such a detector is
underway~\cite{dirdec}, annual-modulation searches are already
feasible using large detector masses~\cite{dama}.

If the local WIMP velocity distribution is isotropic then the annual
modulation is roughly sinusoidal with a maximum in early June (when
the component of the Earth's velocity in the direction of the Sun's
motion is largest) and amplitude of order a few per-cent.  The DAMA
collaboration, using a detector consisting of NaI crystal
scintillators, have detected an annual modulation with roughly these
properties, which they interpret as a WIMP
signal~\cite{dama}. Assuming a standard halo model with a Maxwellian
velocity distribution and circular velocity $v_{{\rm c}}=220 \, \rm km
\, s^{-1}$, they find a best fit WIMP mass $m_{\chi}= 52$ GeV and
cross-section $ \zeta \sigma_{{\rm p}}= 7.2 \times 10^{-9} \, {\rm
nb}$~\footnote{Here $\zeta=\rho_{\chi} / (0.3 \, {\rm GeV \, cm^{-3}})
$ parameterizes the uncertainty in the local WIMP density,
$\rho_{\chi}$.} with the 3-$\sigma$ allowed region encompassing masses
and cross-sections in the range $30 \, {\rm GeV} < m_{\chi} < 100 \,
{\rm GeV}$~\footnote{The lower limit $m_{\chi} > 30$ GeV is imposed by
hand and represents the, somewhat model dependent, limit on the
neutralino mass from accelerator sparticle searches
(e.g.~\cite{acc}).}  and $10^{-9} \, {\rm nb} < \zeta \sigma_{{\rm p}}
< 10^{-8} \, {\rm nb}$~\cite{dama}.

Taken at face value this allowed region is incompatible with the
exclusion limits from the Cryogenic Dark Matter Search
(CDMS)~\cite{CDMS}, Edelweiss~\cite{edelnew} and Zeplin
I~\cite{zeplin} experiments. While these exclusion limits depend
relatively weakly on the WIMP velocity distribution (varying by of
order tens of per-cent for fixed $v_{{\rm c}}$~\cite{rate,myhalomod})
the annual modulation signal depends sensitively on the halo model
assumed~\cite{amdama,amme,amgen,uk,gg,damare,ck,fs} and the region of
WIMP mass--cross-section parameter space corresponding to the DAMA
signal may be significantly enlarged if non-standard halo models are
considered~\cite{amdama,amme,damare}. In particular Belli
et. al.~\cite{damare} carried out an extended analysis of the DAMA
data, for a large range of halo models and parameters, and found a
significant enlargement of the allowed region to encompasses masses in
the range $30 \, {\rm GeV} < m_{\chi} < 270 \, {\rm GeV}$ and
cross-sections in the range $10^{-10} \, {\rm nb} < \zeta \sigma_{{\rm
p}} < 6 \times 10^{-8} \, {\rm nb}$. However, as recently pointed out
by Copi and Krauss~\cite{ck} and Fornengo and Scopel~\cite{fs}, the
phase of the annual modulation can be significantly different if the
WIMP velocity distribution is anisotropic and the time variation may
not be close to sinusoidal~\cite{gg,fs}. In fact, in apparent
contradiction to the results of Belli et. al.~\cite{damare}, Copi and
Krauss~\cite{ck} found, considering a similar range of halo models,
that the DAMA annual modulation signal could not be made compatible
with the exclusion limits from the CDMS and Edelweiss experiments.

Motivated by this apparent conflict, and also the possibility that if
a WIMP annual modulation signal were detected its phase and
shape~\cite{gg,ck,fs} could allow us to probe the structure of the
Milky Way (MW) halo, we study in detail the astrophysical
uncertainties in the calculation of the annual modulation signal. In
Section II we concentrate on the motion of the Earth relative to the
Galactic rest frame. In Section III we discuss the local WIMP velocity
distribution, in particular the importance of using physically and
observationally reasonable models. In Section IV we examine the
effects of the modeling of the Earth's motion and the local velocity
distribution on the phase and amplitude of the signal and the extent
to which it can be approximated by a sinusoidal variation, before
concluding with discussion in Section V.

\section{Motion of the Earth}
\label{analysis}

The differential WIMP elastic scattering rate due to scalar
interactions is given by~\cite{jkg}:
\begin{equation}
\label{drde}
\frac{{\rm d} R}{{\rm d}E}(E, \, t) = \zeta \sigma_{{\rm p}} 
              \left[ \frac{\rho_{0.3}}{\sqrt{\pi} v_{c}}
             \frac{ (m_{{\rm p}}+ m_{\chi})^2}{m_{{\rm p}}^2 m_{\chi}^3}
             A^2 T(E, \, t) F^2(q) \right] \,,
\end{equation}
where $\rho_{0.3} =0.3 \, {\rm GeV \, cm^{-3}}$ (and $\zeta$ is defined
so that the local WIMP density is $\zeta \, \rho_{0.3}$), $\sigma_{{\rm
p}}$ is the WIMP scattering cross section on the proton, $v_{{\rm c}}$
is the local circular velocity, $A$ and $F(q)$ are the mass number and
form factor of the target nuclei respectively, $E$ is the recoil
energy of the detector nucleus, and $T(E)$ is defined, so as to be
dimensionless, as~\cite{jkg}
\begin{equation}
\label{tq}
T(E, \, t)=\frac{\sqrt{\pi} v_{c}}{2} \int^{\infty}_{v_{{\rm min}}} 
            \frac{f_{v}(t)}{v} {\rm d}v \,,
\end{equation}
where $f_{v}(t)$ is the WIMP speed distribution in the rest frame of the
detector, normalized to unity, and $v_{{\rm min}}$ is the minimum
WIMP speed that can cause a recoil of energy $E$:
\begin{equation}
v_{{\rm min}}=\left( \frac{ E (m_{\chi}+m_{A})^2}{2 m_{\chi}^2 m_{A}} 
             \right)^{1/2} \,.
\end{equation}
where $m_{A}$ is the Atomic mass of the detector nuclei.

The WIMP velocity distribution in the rest frame of the detector is
found by making a, time dependent, Galilean transformation ${{\bf v}}
\rightarrow {\tilde{\bf v}}= {\bf v} + {\bf v_{{\rm e}}}(t)$, where ${\bf
v_{{\rm e}}}(t)$ is the Earth's velocity relative to the Galactic rest
frame, which has two components: the Earth's orbit about the Sun and
the Sun's motion with respect to the Galactic rest frame.

\subsection{Orbit about the Sun}

 The Earth moves in a close to circular orbit, inclined at an
angle of roughly $60^{\circ}$ to the Galactic plane, with orbital
speed $v_{{\rm e}} = 29.79 \, {\rm km \, s^{-1}}$. The annual
modulation is mainly determined by the component of the Earth's motion
in the direction of the Sun's orbit~\footnote{For simplicity we ignore
the Sun's motion with respect to the local standard of rest in this
sub-section.}:
\begin{equation}
v_{{\rm e}, \, \odot}(t) = v_{{\odot}} + v_{{\rm e}} \sin{\gamma} \, 
                 \cos{\alpha(t)} \,,
\end{equation}
where $\alpha(t)= 2 \pi (t-t_{{\rm 0}})/ T$, $T=1$ year, $t_{{\rm 0}}
\sim 153$ days (June 2nd), $\gamma \approx 30^{\circ}$ is the angle
between the axis of the ecliptic and the galactic plane and $t$ is
measured in days.  Since $v_{\odot} \gg v_{e}$, if the velocity
distribution is close to isotropic, the differential event rate can be
expanded in a Taylor series in $\cos \alpha(t)$ so that, to first
order,~\cite{amtheory}:
\begin{equation} 
\frac{{\rm d} R}{{\rm d} E_{{\rm }}}(E, \, t) \approx 
         R_{0}(E) 
           + R_{1}(E) \cos{\alpha(t)} \,.
\end{equation}
with $R_{0}(E) \gg R_{1}(E)$, i.e. the modulation is roughly
sinusoidal. 

A commonly used~\cite{amgen,ckdir,amme,ck} expression for the Earth's
motion, in Galactic co-ordinates (X,Y,Z) where X is toward the
Galactic center, Y in the direction of rotation and Z toward the north
Galactic pole, is:
\begin{equation}
\label{ve1}
{\bf v_{{\rm e}}}(t) = v_{{\rm e}} \left[ -\sin{\alpha(t)},  
             \,  \cos{\alpha(t)} \sin{\gamma} ,  \,
                  -\cos{\alpha(t)} \cos{\gamma} \right]  \,.
\end{equation}
This expression assumes that the
Earth's orbit is circular and the axis of the ecliptic lies in the Y-Z
plane~\cite{amgen}.

A more accurate expression for the components of the Earth's
velocity relative to the Sun can be found by using the expression for
the Sun's ecliptic longitude (from e.g.~p77 of Ref.~\cite{ref}),
$\lambda= L + 1^{\circ}.915 \sin{g} + 0.020^{\circ} \sin{2g}$ where
$L=280^{\circ}.460 + 0^{\circ}.9856003 \,t$ is the mean longitude of
the sun, corrected for aberration, $g=357^{\circ}.528 +
0^{\circ}.9856003\ t$ is the mean anomaly (polar angle of orbit) and $t$
is the time in days from 1200 Universal Time~\footnote{Universal Time
is equivalent to Greenwich Mean Time to within an accuracy of
seconds.} on Dec 31st 1999, and transforming from ecliptic to Galactic
co-ordinates (e.g. p13 of Ref.~\cite{ref}). Lewin and Smith~\cite{ls}
carried out his procedure and found:
\begin{eqnarray}
\label{vels}
{\bf v_{{\rm e}}}(t)& = & v_{{\rm e}}(\lambda) \left[ 
       \cos{\beta_{x}} \sin{ \left( \lambda - \lambda_{x} \right) }, 
         \right. \,  
       \nonumber \\
      && \left.  \cos{\beta_{y}} \sin{ \left( \lambda - \lambda_{y} \right)} ,
     \,\,  \cos{\beta_{z}} \sin{ \left( \lambda - \lambda_{z} \right)} \right] \,,
\end{eqnarray}
where $v_{{\rm e}}(\lambda) = v_{{\rm e}} \left[ 1 - e \sin{(\lambda -
\lambda_{0})} \right]$, $e=0.016722$ and $\lambda_{0}= 13^{\circ} \pm
1^{\circ}$ are the ellipticity of the Earth's orbit and the ecliptic
longitude of the orbit's minor axis respectively, and $\beta_{i}=
(-5^{\circ}.5303, 59^{\circ}.575, 29^{\circ}.812)$ and $\lambda_{i}=
(266^{\circ}.141,-13^{\circ}.3485,179 ^{\circ}.3212)$ are the ecliptic
latitudes and longitudes of the of the (X,Y,Z) axes respectively.

\begin{figure}[t]
\centering
\includegraphics[width=0.45\textwidth]{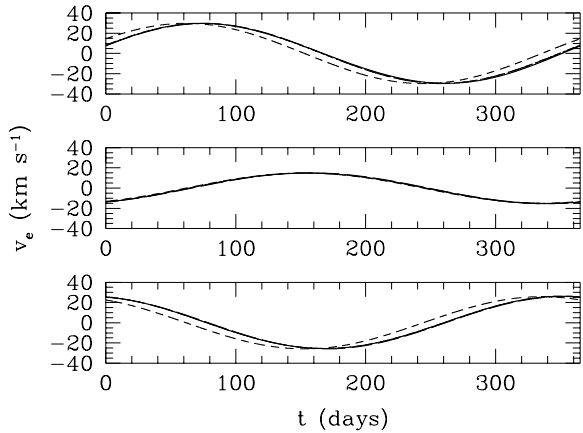}
\caption[fig1]{\label{fig1} The components of the orbital velocity of
the Earth in Galactic X,Y,Z co-ordinates ($v_{{\rm e, \, X}}$ top
panel, $v_{{\rm e, \, Y}}$ middle panel and $v_{{\rm e, \, Z}}$ bottom
panel) found using the assumptions discussed in the text: i) assuming
that the Earth's orbit is circular and the ecliptic lies in the X-Y
plane (eq.~(\ref{ve1}), short dashed lines) ii) ignoring the
ellipticity of the Earth's orbit and the non-uniform motion of the Sun
in right ascension (eq.~(\ref{vegg}) and Ref.~\cite{gg}, long dashed)
iii) including the ellipticity of the Earth's orbit but not the
non-uniform motion of the Sun (Ref.~\cite{fs}, dot-dashed) and iv)
including the ellipticity of the Earth's orbit and the non-uniform
motion of the Sun (eq.~(\ref{vels}) and Ref.~\cite{ls}, solid). Time
is measured in days from noon on Dec 31st 2002.}
\end{figure}

Fornengo and Scopel~\cite{fs} used a similar expression where the the
non-uniform motion of the Sun in right ascension is neglected in the
co-ordinate transformation so that the $\sin{(\lambda-\lambda_{i})}$
terms in eq.~(\ref{vels}) above are replaced by $\cos{\left[ \omega
(t-t_{i}) \right]}$ where $t_{i}=(76.1, 156.3, 352.4)$ days and
$\omega= 2 \pi /(1 \, {\rm year})$. Gelmini and Gondolo~\cite{gg} used
a slightly simpler expression, which neglects the ellipticity of the
Earth's orbit and the non-uniform motion of the Sun in right
ascension. In terms of the Sun's ecliptic longitude, which under these
approximations is given by $\tilde{\lambda}(t)= \omega (t-t_{1})$
where $t_{1}$ is the spring (vernal) equinox, $t_{1}=79.55$
days~\footnote{This figure is for 2003 and is accurate to within about
0.04 days. The time at which the Spring equinox occurs increases by
0.24 days a year (see e.g. Ref.~\cite{usno}).},:
\begin{equation}
\label{vegg}
{\bf v_{{\rm e}}}(t) = v_{{\rm e}} \left[ \hat{{\bf e}}_{1}
\sin{\tilde{\lambda}(t)} - \hat{{\bf e}}_{2} \cos{\tilde{\lambda}(t)}
\right] \,,
\end{equation}
where $\hat{{\bf e}}_{1}$ and $\hat{{\bf e}}_{1}$ are in the Sun's
direction at the Spring equinox and Summer solstice respectively:
\begin{eqnarray}
 \hat{{\bf e}}_{1} &=& (-0.0670, \, 0.4927, \, -0.8676) \,, \nonumber \\
 \hat{{\bf e}}_{2} &=& (-0.9931, \, -0.1170, \, 0.01032)  \,.
\end{eqnarray}

These expressions for the components of the Earth's orbital velocity
are displayed in Fig.~1, and the deviations between the expressions in
Fig.~2. In the plots time is measured in days from noon on Dec 31st
2002, but the correct time (measured in days from noon on Dec 31st
1999) is used in the calculations. If time is erroneously measured
from the beginning of the year rather than the beginning of 2000, then
this leads to a horizontal shift in the curves of 0.27 days for every
year elapsed since 2000. The deviations in the Y component are
reassuring small (of order a few per-cent), however assuming that the
Earth's orbit is circular and the axis of the ecliptic lies in the Y-Z
plane (eq.~(\ref{ve1})) leads to large errors in the X and Z
components. The other three expressions are in relatively good
agreement for all three components.

\begin{figure}[t]
\centering
\includegraphics[width=0.45\textwidth]{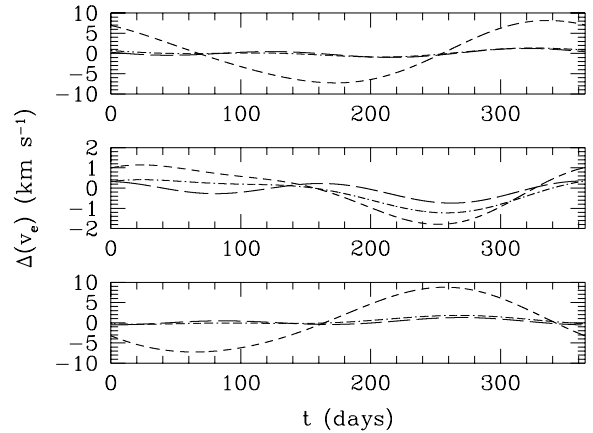}
\caption[fig2]{\label{fig2} The differences between the expressions
for the velocity of the Earth in Galactic co-ordinates ($ \Delta
(v_{{\rm e, \, X}}), \, \Delta (v_{{\rm e, \, Y}}), \, \Delta (v_{{\rm e, \,
Z}}) $, top, middle and bottom panel respectively) relative to the full
expression including the ellipticity of the Earth's orbit and the
non-uniform motion of the Sun (eq.~(\ref{vels}) and Ref.~\cite{fs}).
Line types as in Fig. 1. Note the larger scale used for $\Delta
(v_{{\rm e, \, Y}})$.}
\end{figure}

\subsection{Motion of the Sun}

We now consider the motion of the Sun, which can be divided into two
components: the motion of the local standard of rest (LSR) and the
Sun's peculiar motion with respect to the LSR, $v_{\odot, \, {\rm
pec}}$. The ``standard values'' often used in WIMP event rate
calculations date back to Kerr and Lynden-Bell's 1986 ``Review of
Galactic constants''~\cite{stand}, however since then the Hipparcos
satellite has provided accurate measurements of the motions of large
numbers of nearby stars, allowing more accurate determinations of the
Galactic constants, so the ``standard values'' should be updated.

The standard value used for the Sun's peculiar motion, with respect to
the LSR, is $v_{\odot, \, {\rm pec}} = (10, 15, 8) \, {\rm km \,
s^{-1}}$~\cite{stand}, while Lewin and Smith~\cite{ls} use $v_{\odot,
\, {\rm pec}} = (9, 12, 7) \, {\rm km \, s^{-1}}$, with errors of
order a few ${\rm km \, s^{-1}}$ in each component leading to an
uncertainty of several days in the phase of the
modulation~\cite{gg}. The value determined more recently using data
from Hipparcos is $v_{\odot, \, {\rm pec}} = (10.0, 5.2, 7.2) \, {\rm
km \, s^{-1}}$~\cite{hipp,bm}, with the stated errors in each
component being roughly $0.5 \, {\rm km \, s^{-1}}$.

Assuming that the MW is axisymmetric, then the motion of the LSR is,
by definition, $(0,v_{{\rm c}}(R_{0}),0)$ where $v_{{\rm c}}(R_{0})$
is the circular velocity at the solar radius. Kerr and Lynden-Bell
combined a large number of independent determinations of the circular
velocity and found $v_{{\rm c}}= 222.2 \, {\rm km \, s^{-1}}$, with
standard deviation $20 \, {\rm km \, s^{-1}}$~\cite{stand}. The proper
motions of Cepheids measured by Hipparcos allow accurate
determinations of the Oort constants $A$ and $B$
~\cite{oort}, which lead to a value for the circular velocity, in
terms of the Galactocentric distance $R_{0}$: $v_{{\rm c}}(R_{0})=
(27.2 \pm 0.9) (R_{0}/ {\rm kpc}) \, {\rm km \,
s^{-1}}$~\cite{bm}. One possible caution is that this calculation
assumes that stars move on circular orbits, which may lead to a
systematic error as the MW is not exactly
axisymmetric~\cite{bm}. Accurate very long baseline interferometry
measurements of the proper motion of the ${\rm Sgr A}^{*}$ radio
source at the Galactic center provides an alternative, marginally
inconsistent, determination of $v_{{\rm c}}$: $(30.1 \pm 0.9) (R_{0}/
{\rm kpc}) {\rm km \, s^{-1}}$~\cite{sgra,bm}. In this case non-zero
peculiar motion of ${\rm Sgr A}^{*}$ produces a systematic error.
Using $R_{0}=8.0 \pm 0.5 \, {\rm kpc}$ from the most recent
compilation of determinations of $R_{0}$~\cite{R0}, leads to $v_{{\rm
c}}(R_{0})= (218 \pm 15) \, {\rm km \, s^{-1}}$ from Cepheids and
$v_{{\rm c}}(R_{0})= (241 \pm 17) \, {\rm km \, s^{-1}}$ from ${\rm
Sgr A}^{*}$.

\section{Local WIMP velocity distribution}
\label{halodmod}

Data analyzes nearly always assume a standard halo model with an
isotropic Maxwellian velocity distribution:
\begin{equation}
\label{max}
f({\bf v}) = \frac{1}{ \pi^{3/2} \, v_{{\rm c, \, h}}^3} \exp{ (- 
         {\bf v}^2 / v_{{\rm c, h}}^2)} \,,
\end{equation}
where $v_{{\rm c, \, h}}$ is the contribution of the halo to the local
circular velocity, which corresponds to a spherically symmetric
density distribution with $\rho \propto r^{-2}$ i.e. an isothermal
sphere. Observations and numerical simulations indicate that galaxy
halos are in fact triaxial and anisotropic however (see
Ref.~\cite{myhalomod} for a review). As the
size~\cite{amdama,amgen,uk,gg,damare,ck,fs} and phase~\cite{ck,fs} of
the annual modulation signal depend quite sensitively on the halo
model assumed, the realistic modeling of the WIMP velocity
distribution is crucial when extracting parameters or exclusion limits
from data or comparing results from different experiments.

The steady state phase-space distribution function of a collection of
collisionless particles is given by the collisionless Boltzmann
equation, and the velocity dispersions, $<v_{{\rm i}}^2>$, of the
system are calculated via the Jeans equations, which are found by
taking moments of the collisionless Boltzmann equation (see
e.g.~\cite{bt}). Solutions to the Jean's equations have the property
that the tensor $\sigma_{{\rm ij}}^2 \equiv \overline{ (v_{{\rm i}} -
\bar{v}_{{\rm i}}) (v_{{\rm j}} - \bar{v}_{{\rm j}})}$ is symmetric,
so that at any point a set of orthogonal axes can by chosen such that
${\bf{ \sigma}}$ is diagonal (i.e. $\sigma_{{\rm ij}}^2 = \sigma_{{\rm
ii}}^2 \delta_{{\rm ij}}$)~\cite{bt}. However, as discussed in
Refs.~\cite{bt,newevans}, there is no equation of state relating the
components of the velocity dispersion to the density, so solving the
Jeans equations requires assumptions, which may or may not be
physically reasonable, about the shape and/or orientation of $\mathbf
{\sigma}$. Evans et. al.~\cite{newevans} presented the logarithmic
ellipsoidal model (which has potential $\Phi(x,y,z)= (v_{{\rm c}}^2/2)
\ln{( x^2 + y^2 p^{-2}+ z^2 q^{-2})}$) which is the simplest triaxial,
scale free generalization of the isothermal sphere. They argued that
principle axes aligned with conical co-ordinates correspond to
physical distribution functions and calculated the corresponding
velocity dispersions. On the axes of the halo conical co-ordinates are
locally equivalent to cylindrical polar co-ordinates and the local
velocity distribution can be approximated by a multi-variate Gaussian
with principle axes aligned with the (X,Y,Z) Galactic axes. This form
has been widely used in calculations of the WIMP direction detection
rate~\cite{newevans,ckdir,damare,myhalomod,ck,fs}, however a
multi-variate Gaussian with arbitrary velocity dispersions need not
correspond to a physically sensible halo model. For instance Evans
et. al. do not consider parameter values for which the ratio of any
two of the velocity dispersions is more than 3:1, so as to avoid
models which are afflicted by instabilities~\cite{newevans}.

We will now briefly examine what sets of values of the velocity
dispersions are reasonable for the logarithmic ellipsoidal model, if
the Sun is located on the intermediate axis of the halo (the
deviations from an isotropic velocity distribution are smaller on the
major axis~\cite{newevans,myhalomod}). On the intermediate axis the
velocity distribution can be written as
\begin{equation}
f({\bf v})= \frac{1}{ (2 \pi)^{3/2}  \sigma_{{\rm r}}  \sigma_{{\phi}} 
      \sigma_{{\rm z}}} {\rm exp} \left( - 
            \frac{v_{{\rm r}}^2}{2 \sigma_{{\rm r}}^2} - 
           \frac{v_{{\phi}}^2}{2 \sigma_{{\phi}}^2} -
            \frac{v_{{\rm z}}^2}{2 \sigma_{{\rm z}}^2} \right)  \,,
\end{equation}
with the velocity dispersions given by
\begin{eqnarray}
 \sigma_{{\rm r}}^2 &=& \frac{v_{{\rm c, \, h}}^2 \,\, p^{-4}} {(2+
            \gamma)(1-p^{-2} + q^{-2})} \,, \nonumber \\ 
\sigma_{{\phi
            }}^2 &=& \frac{v_{{\rm c, \, h}}^2 \, ( 2 q^{-2}
            -p^{-2})}{2(1-p^{-2} + q^{-2})} \,, \nonumber \\
\sigma_{{\rm z}}^2 &=& \frac{v_{{\rm c, \, h}}^2 \, (2-
            p^{-2}) } {2(1-p^{-2} + q^{-2})} \,,
\end{eqnarray} 
where $p$ and $q$ are constants which satisfy $ 0 \leq q \leq
p \leq 1$ and are related to the axial ratios of the density
distribution, $I_{1,2}$, by
\begin{eqnarray}
I_{1}^2 & = & \frac{ p^2 \, (p^2 q^2 +p^2 -q^2)}{q^2 + p^2 - p^2 q^2} \,,
        \nonumber \\
I_{2}^2 & = & \frac{ q^2 \, (p^2 q^2 -p^2 +q^2)}{q^2 + p^2 - p^2 q^2} \,,
\end{eqnarray}
and $\gamma$ is a constant isotropy parameter, which in the spherical
limit $p=q=1$ is related to the anisotropy parameter 
\begin{equation}
\beta=1 - \frac{<v_{\theta}^2>+<v_{ \phi}^2>}{2 <v_{{\rm r}}^2>} \,,
\end{equation}
by $-\gamma= 2 \beta$. Note that $\sigma_{{\phi }}^2 + \sigma_{{\rm z
}}^2= v_{{\rm c, \, h}}^2$ and $\sigma_{{\phi }} > \sigma_{{\rm z }}$. 

If we require $\sigma_{{\rm j}}/3 < \sigma_{{\rm i}} < 3 \sigma_{{\rm
j}}$, so as to avoid models which might be affected by
instabilities~\cite{newevans}, then the velocity dispersions lie in
the ranges $0.32 < \sigma_{{\rm z}}/v_{{\rm c, \, h}} < 0.71$, $0.71 <
\sigma_{{\phi}}/v_{{\rm c, \, h}} < 0.95$ and $0.24 < \sigma_{{\rm
r}}/v_{{\rm c, \, h}} < 2.0$. If we also require that $ 0 < \beta <
0.4$, as found at the solar radius in simulated halos~\cite{mooredm},
then $\sigma_{{\rm r}}$ is further restricted to $0.71 < \sigma_{{\rm
r}}/v_{{\rm c, \, h}} < 0.92$. In Fig. 3 we plot the values of
$\sigma_{{\rm r}}$ and $\sigma_{{\phi }}$ ($\sigma_{{\rm z
}}^2=v_{{\rm c, \, h}}^2- \sigma_{{\phi }}^2$) which satisfy these two
requirements. We still need to check that the axial ratios of the
density distribution are in reasonable agreement with observed and
simulated galaxy halos. There is no simple algebraic relation between
the $ I_{{\rm i}}$ s and the $\sigma_{{\rm i }}$ s. If we require $0.6
< I_{1} < 1.0 $ and $0.3 < I_{2} < 1.0$ (see e.g. Ref.~\cite{sackett}
for a review of constraints on the shape of dark matter halos) then
some of the previously allowed sets of velocity dispersions are now
excluded.

The logarithmic ellipsoidal model may not be a unique solution to the
Jeans equations (and there is no reason to expect the Sun to be
located on one of the axes of the halo), but otherwise there is no
reason for the principle axes of the velocity distribution to
correspond to the axes of the halo. In other words a multivariate
Gaussian velocity distribution with axes corresponding to the axes of
the halo and arbitrary velocity dispersions may not correspond to a
physically reasonable halo model. While the constraints on the axial
ratios and anisotropy of the halo we have imposed are not cast in
stone, they illustrate that only restricted sets of values of the
velocity dispersions correspond to observationally and physically
reasonable halo models, and that the ratios of the velocity
dispersions can not be too large. Simulations provide further support
for this argument. Helmi, White and Springel~\cite{hws} examined the
simulation particles within a 4 $\, {\rm kpc}$ box located 8 $\, {\rm
kpc}$ from the center of a Milky Way like halo and found that, apart
from a clump of fast moving particles from a late accreting subhalo,
the velocity distribution was well approximated by a multi-variate
Gaussian, with principal axis velocity dispersions in the ratio
$1:1.08:1.27$. If in fact the local dark matter distribution is not
smooth (see below for discussion of this possibility), then the local
velocity distribution could not be approximated by a multivariate
Gaussian with any set of velocity dispersions. We conclude that the
large velocity dispersion ratios found by Fornengo and Scopel to
produce extreme distortions of the annual modulation ($(1:5:4)$ and
$(10:1:3)$)~\cite{fs} are unlikely to correspond to a realistic halo
model.

\begin{figure}[t]
\centering \includegraphics[width=0.45\textwidth]{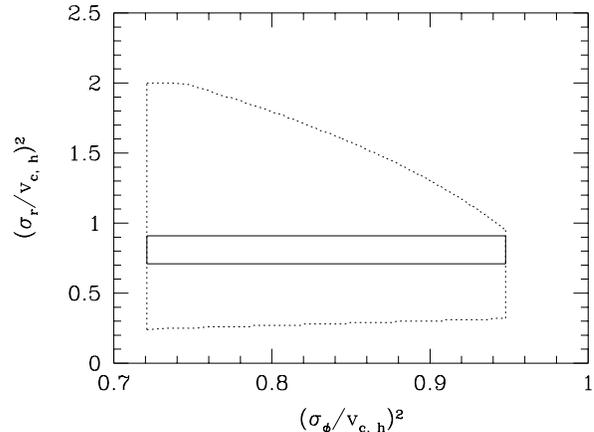}
\caption[fig3]{\label{fig3} Physically and observationally reasonable
values of $\sigma_{{\rm r}}$ and $\sigma_{{\phi }}$ ($\sigma_{{\rm z
}}=(v_{{\rm c, \, h}}^2- \sigma_{{\phi }}^2)^{1/2}$). Inside the dotted
lines the ratio of any two of the velocity dispersions is no greater
than 1:3 and inside the solid lines the anisotropy parameter $\beta$
is in the range $ 0 < \beta < 0.4$. Restricting the axial ratios of
the density distribution would further rule out some sets of values
inside the solid lines.}
\end{figure}

\vspace{1.0cm}

It is possible that the local WIMP distribution may not be completely
smooth, or in other words that the phase-space distribution has not
reached a steady state.  The density and velocity distributions of the
``particles'' in numerically simulated halos are relatively smooth at
the solar radius~\cite{mooredm,hws}, however WIMP direct detection
experiments probe the dark matter distribution within a local sub-mpc
region and even the highest resolution simulations have less than 100
``particles'' per ${\rm kpc}^3$. Furthermore the first neutralino
clumps to form have mass more than ten orders of magnitude smaller
than the smallest subhalos resolved in numerical
simulations~\cite{hss}, and it is possible that the cores of some of
these first small, high density, clumps could survive to the present
day~\cite{hss,myhalomod,ber}. Stiff and Widrow~\cite{swnew} have
recently developed a technique which uses test particles to probe the
velocity distribution at a single point within a numerical
simulation. They find a local velocity distribution consisting of a
series of discrete peaks, and argue that the smooth background found
previously in simulations may be, at least partly, a consequence of
the use of a finite volume to measure the velocity distribution. Even
if the local phase-space distribution bears no trace of the first WIMP
clumps to form, more massive subhalos which have been accreted onto
the MW relatively recently, and have sufficiently eccentric orbits,
could produce coherent streams of high velocity WIMPs at the
solar radius~\cite{swf,hws}.

\vspace{1.0cm}

An additional complication is that the contribution of the visible
components of the MW to the circular velocity at the Solar radius is
non-negligible, and may even dominate that of the halo (see
e.g. Refs.~\cite{massmod,mooredm}). Self consistent halo models can in
principle be built (taking into account various observations) using
Eddington's formula~\cite{uk,damare}, however building mass models of
the Milky Way is a complex and ill-constrained
process~\cite{massmod,bm}; a small change in the observational
constraints can lead to a large change in the properties of the dark
halo. In Dehnen and Binney's mass models, where the halo is described
by a spheroidal density distribution, the contribution of the MW halo
to the circular velocity at the solar radius lies in the range $
v_{{\rm c, \, h}}(R_{0})=110-180 \, {\rm km \,
s^{-1}}$~\cite{massmod}, while Moore et. al. find, for halos with a
central density cusp $\rho \propto r^{-1.5}$, taking $ 190 \, {\rm km
\, s^{-1}} < v_{{\rm c}}(R_{0})< 230 \, {\rm km \, s^{-1}}$ and using
a smaller set of constraints, $ 100 \, {\rm km \, s^{-1}} < v_{{\rm c,
\, h}}(R_{0}) < 120 \, {\rm km \, s^{-1}}$~\cite{mooredm}. We
therefore conclude that a relatively conservative range of values is:
$100 \, {\rm km \, s^{-1}} < v_{{\rm c, \, h}}(R_{0}) < 200 \, {\rm km
\, s^{-1}}$, and that the ``standard'' value, $v_{{\rm c, \,
h}}(R_{0})= 220 \, {\rm km \, s^{-1}}$, is probably too large.

\section{Results}

\begin{figure}[t]
\centering
\includegraphics[width=0.45\textwidth]{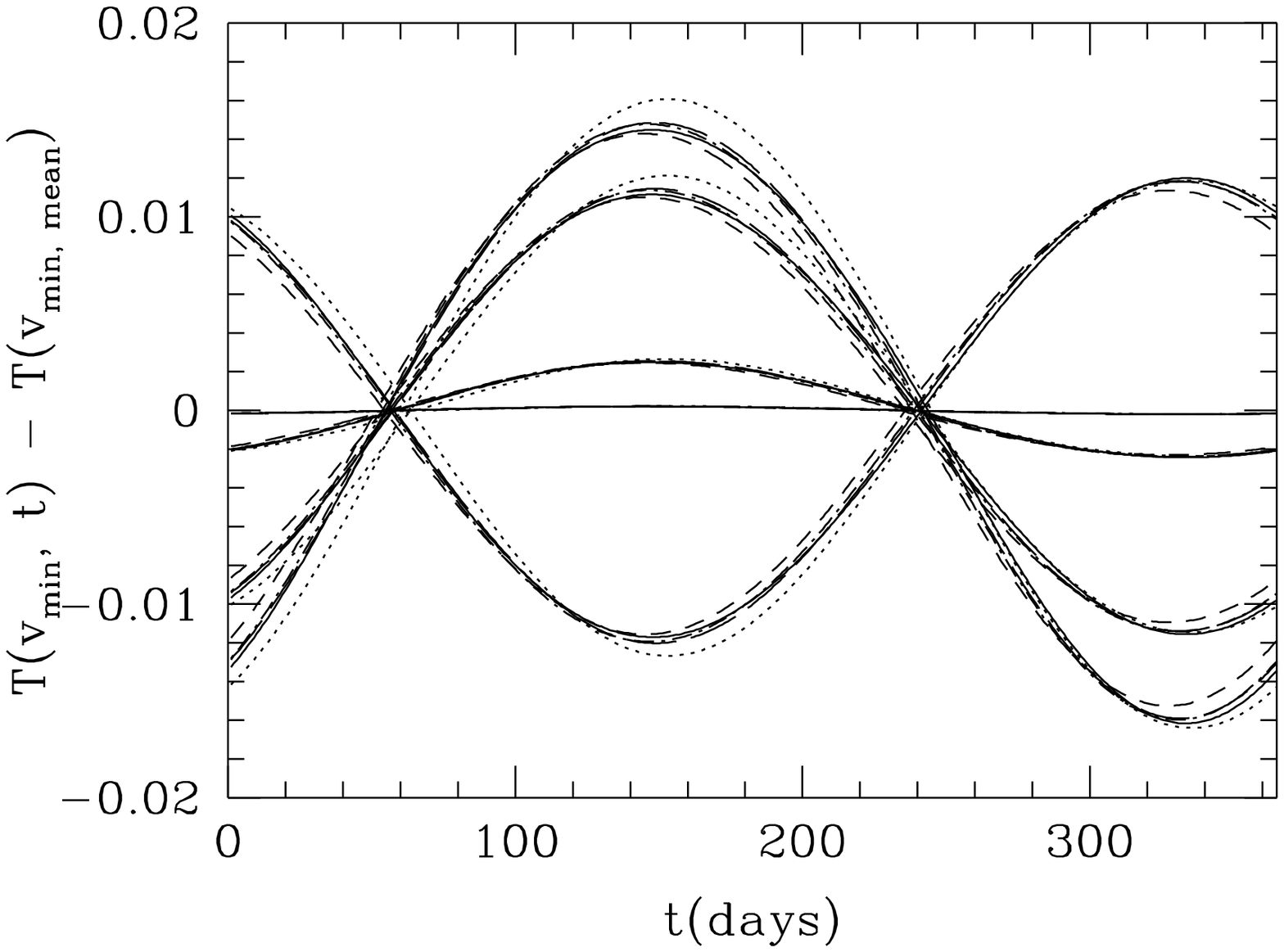}
\includegraphics[width=0.45\textwidth]{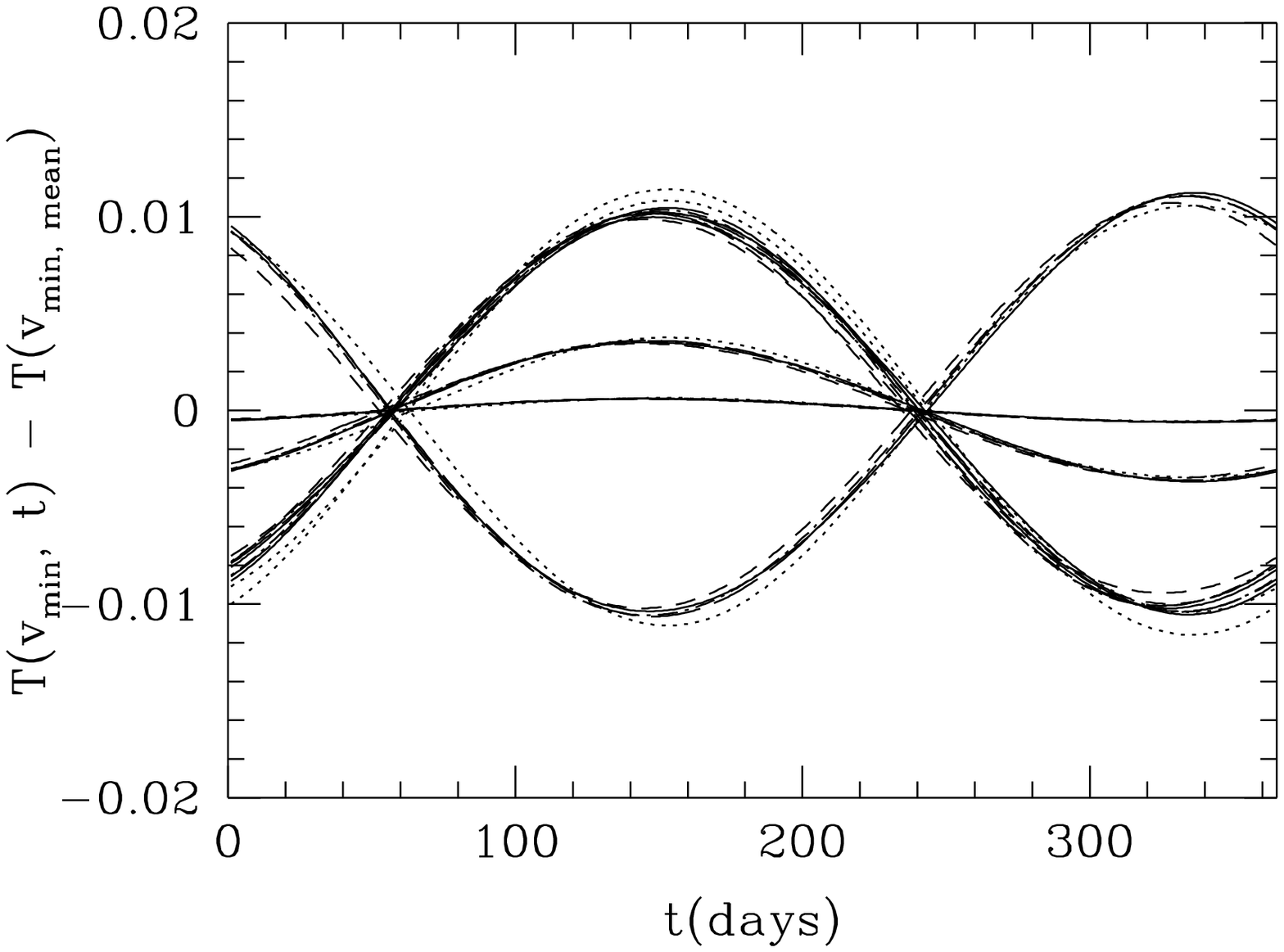}
\caption[fig4]{\label{fig4} The dependence of $T(v_{{\rm min}}, \, t)$
on the approximations used when calculating the Earth's orbit for,
from top to bottom at t=0, $v_{{\rm min}}=100, 500, 400, 300$ and $200
\, {\rm km s^{-1}}$ with line types as in Fig.~1, for the standard
halo model (upper panel) and for the fiducial triaxial model, see text
for details, (bottom panel). Note the change in the phase of the
annual modulation for $v_{{\rm min}}=100 \, {\rm km s^{-1}}$ (see text
and Ref.~\cite{revphase} for further discussion of the change in phase
which occurs for small $v_{{\rm min}}$). We have
fixed $v_{{\rm c, \, h}} = 150 \, {\rm km \, s^{-1}}$ and $v_{\odot,
\, {\rm pec}} = (10.0, 5.2, 7.2) \, {\rm km \, s^{-1}}$ here.}
\end{figure}

\begin{figure}[t]
\centering
\includegraphics[width=0.45\textwidth]{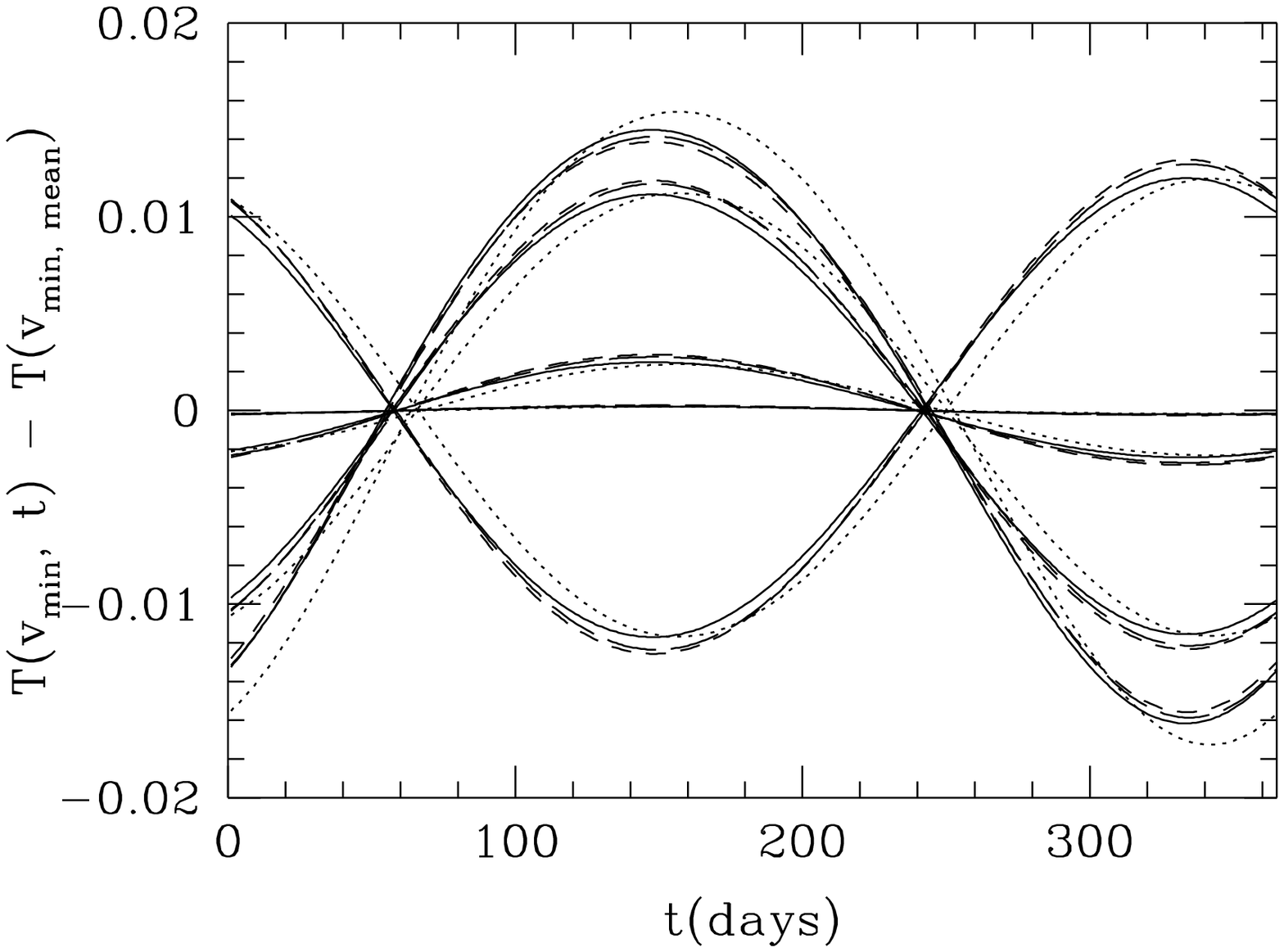}
\includegraphics[width=0.45\textwidth]{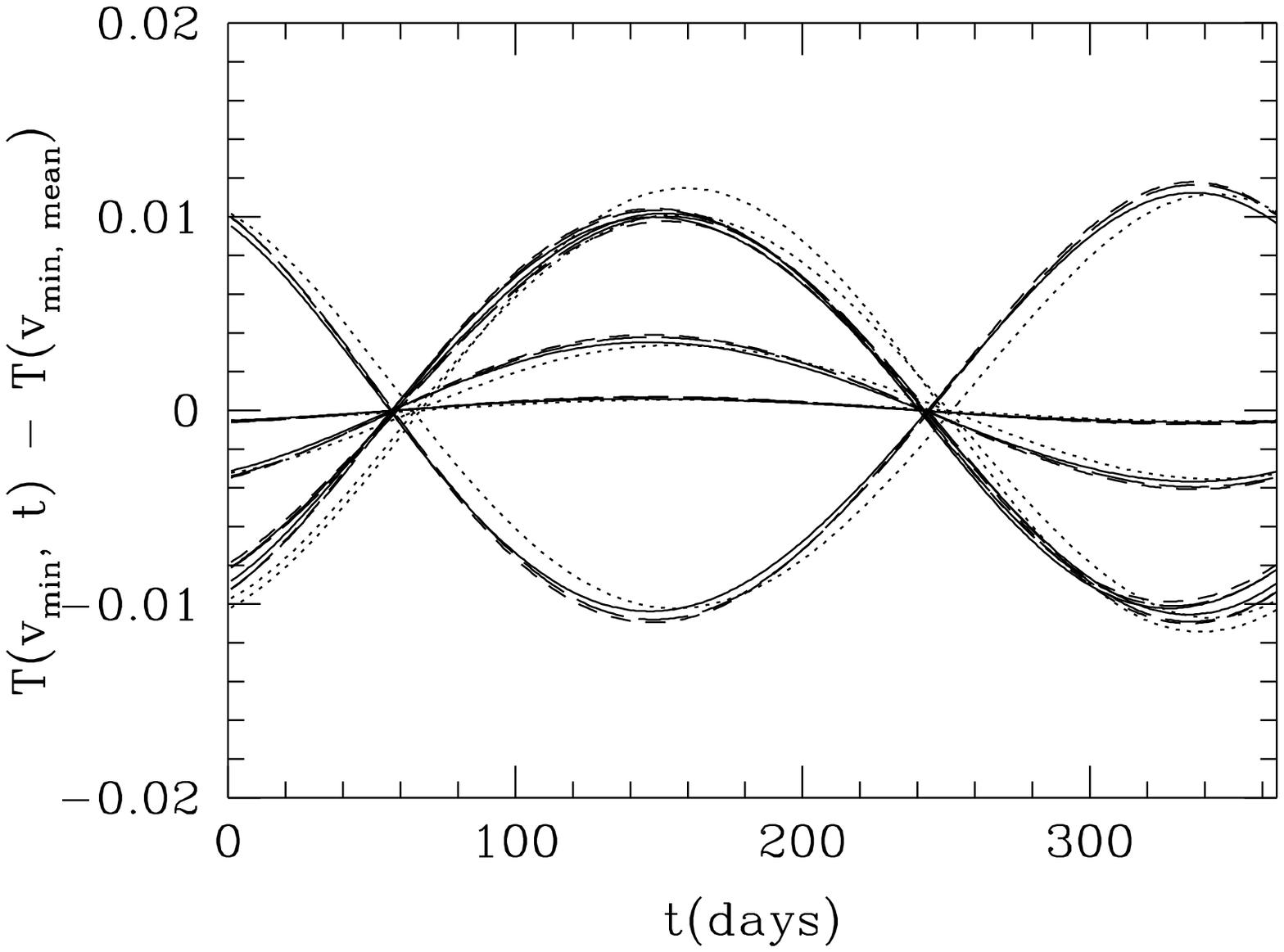}
\caption[fig5]{\label{fig5} As Fig.~4 for the dependence of $T(v_{{\rm
min}}, \, t)$ on the values used for the Sun's velocity with respect to
the LSR ($v_{\odot, \, {\rm pec}}$, in ${\rm km \, s^{-1}}$): (10.0, 5.2,
7.2) (Ref.~\cite{hipp,bm}, solid line), (0, 0, 0) (dotted line), (10,
15, 8) (Ref.~\cite{stand}, short dashed), (9, 12, 7) (Ref.~\cite{ls},
long dashed). The full expression for the Earth's orbit (Ref.\cite{ls}
and eq.~\ref{vels}) is used here.}
\end{figure}

To compare theoretical predictions with experimental data requires the
detector atomic mass, form factor and resolution, and the relationship
between the nuclear recoil energy and the energy detected, to be taken
into account (see e.g. Ref.~\cite{ls}). As the DAMA collaboration's
raw data is not publicly available instead of working with the
differential event rate we look at the detector independent quantity
$T(v_{{\rm min}}, \, t)$, as defined in eq.~(\ref{tq}). The relation
between $v_{{\rm min}}$ and the detected energy is given, for various
target nuclei and WIMP masses, in Fig.~4 of Ref.~\cite{fs}.  For
reference, the DAMA experiment has a measured energy threshold of 2
keV which corresponds (for Iodine which dominates) to a recoil energy
threshold of 22 keV, which for $m_{\chi} = 50 \, (200) $ GeV requires
a minimum WIMP velocity $v_{{\rm min}}= 309 \, (146) \, {\rm km \,
s^{-1}}$.

\subsection{Motion of the Earth}

We will start by examining the effects of the modeling of the motion
of the Earth on the phase and shape of the annual modulation
signal. We use two fiducial halo models: the standard halo model (with
an isotropic Maxwellian velocity distribution) and the logarithmic
ellipsoidal model~\cite{newevans}, with the Earth located on the
intermediate axis and parameter values $p= 0.9, q=0.8$ and
$\gamma=-0.62 $ corresponding to axial ratios $1:0.78:0.48$ and $\beta
=0.4$. The later model is intended as a specific, somewhat arbitrary
but not unreasonable, example of an anisotropic halo model. In both
cases we take $v_{{\rm c, \, h}}=150 \, {\rm km \, s^{-1}}$. In Fig.~4
we plot $T(v_{{\rm min}}, \, t)$~\footnote{Our plots look
superficially different to those of Fornengo and Scopel as they
normalize to the event rate on January 1st whereas we subtract the
mean event rate.}  produced by each of the expressions for the Earth's
motion discussed in Sec.~\ref{analysis}, for five specific values of
$v_{{\rm min}}$. As has long been known (see
e.g. Ref.~\cite{revphase}), for small $v_{{\rm min}}$ the phase of the
modulation is reversed. Furthermore as $v_{{\rm min}}$ is increased
the amplitude of the modulation reaches a maximum before decreasing
again. For the standard halo model this maximum occurs at roughly $220
\, {\rm km \, s^{-1}}$, whereas for the fiducial triaxial model the
maximum is at around $250 \, {\rm km \, s^{-1}}$ and $T(\, 200 \, {\rm
km \, s^{-1}}, \, t) \sim T(\, 300\, {\rm km \, s^{-1}}, \, t)$. In
Fig.~5 we plot $T(v_{{\rm min}}, \, t)$ produced by each of the sets
of values for the Sun's motion with respect to the LSR discussed in
Sec.~\ref{analysis}.

\begin{table}
\label{tab1}
\begin{center}
\begin{tabular}{|c|c|c|c|}
Earth's orbit & $t_{{\rm peak}}$ (days) & $\Delta_{{\rm max}}^{{\rm
LS}}$ & $\Delta_{{\rm max}}^{{\rm sin}}$ \\ \hline \hline {\rm
zero-th order} & 153 & -2.4\% -- -15.6\% & 0.1\% -- -11.1\% \\ \hline 
{\rm first order} & 145 & 2.6\% -- 19.0\% & 0.3\% -- -6.6\% \\ \hline 
{\rm GG} & 148 -- 149 & 2.2\% -- 15.6\% & 0.4\% -- -7.2\% \\ \hline 
{\rm FS} & 146 -- 147 & 2.2\% -- 15.8\% & 0.5\% -- -6.7\% \\ \hline 
{\rm LS} & 147 -- 148 & n/a & 0.7\% -- -5.1\% \\
\end{tabular}
\end{center}
\caption[1]{\label{1} The time, $t_{{\rm p}}$, at which $T(v_{{\rm
min}}, \, t)$ is largest, the largest percentage deviation from
$T(v_{{\rm min}}, \, t)$ calculated using the full expression for the
Earth's velocity, $\Delta_{{\rm max}}^{{\rm LS}}$, the largest
percentage deviation from a sinusoidally varying $T(v_{{\rm min}}, \,
t)$ with the same mean, amplitude and $t_{{\rm p}}$, $\Delta_{{\rm
max}}^{{\rm sin}}$, for each of the expressions for the Earth's
velocity discussed in Sec.~\ref{analysis}, for the standard isotropic
halo model. Using only the component of the Earth's motion in the
Galactic plane is denoted by ``zero-th order'', ``first order''
denotes assuming the axis of the ecliptic lies in the Y-Z plane
(eq.~\ref{ve1}), ``GG'' neglecting the ellipticity of the Earth's
orbit and the non-uniform motion of the Sun (Ref.\cite{gg} and
eq.~(\ref{vegg})), ``FS'' neglecting the non-uniform motion of the Sun
(Ref.\cite{fs}) and ``LS'' including the ellipticity of the Earth's
orbit and the non-uniform motion of the Sun (Ref.\cite{ls} and
eq.~(\ref{vels})).  Where a range of values are given the quantity
varies with $v_{{\rm min}}$ and the left (right) hand value is for
$v_{{\rm min}}= 200 \,\, (500) \, {\rm km \, s^{-1}}$. Here we have
fixed $v_{{\rm c, \, h}}=150 \, {\rm km \, s^{-1}}$ and used
$v_{\odot, \, {\rm pec}}= (10.0, 5.2, 7.2) \, {\rm km \, s^{-1}}$.}
\end{table}

\begin{table}
\label{tab2}
\begin{center}
\begin{tabular}{|c|c|c|c|}
Earth's orbit & $t_{{\rm p}}$ (days) & $\Delta_{{\rm max}}^{{\rm
LS}}$ & $\Delta_{{\rm max}}^{{\rm sin}}$ \\ \hline \hline 
{\rm zero-th order} & 153 & -1.5\% -- -11.6\%  & 0.1\% -- -4.7\% \\ \hline
{\rm first order} & 147 -- 142   & 2.2\% -- 12.5\%  & 0.3\% -- -1.0\%  \\ \hline
{\rm GG} & 153 -- 146  & 2.0\% -- 9.6\%  & 0.4\% -- -2.3\% \\ \hline
 {\rm FS} & 151 -- 144  & 2.0\% -- 9.9\%  &  0.4\% -- -2.7\% \\ \hline
{\rm LS}  & 152 -- 144  & n/a  & 0.3\% -- -3.2\%  \\
\end{tabular}
\end{center}
\caption[2]{\label{2} As Table I for the fiducial triaxial halo
model.}
\end{table}

\begin{table}
\label{tab3}
\begin{center}
\begin{tabular}{|c|c|c|c|}
$v_{\odot, \, {\rm pec}} ({\rm km \, s^{-1}}$) & $t_{{\rm p}}$ (days) &
$\Delta_{{\rm max}}^{{\rm H}}$ & $\Delta_{{\rm max}}^{{\rm sin}}$
\\ \hline (0, 0, 0) & 157 & 3.1\% -- 19.3\% & 0.7\% -- 5.6\% \\ \hline 
 (10, 15, 8)  & 148 & 4.0\% -- 30.0\%  & 0.6\% -- 4.9\%  \\ \hline 
  (9, 12, 7)  & 149 & 2.7\% -- 20.0\% & 0.6\% -- 4.9\%   \\ \hline 
 (10.0, 5.2, 7.2)  & 147 -- 148 & n/a & 0.7\% -- 5.1\% \\  
\end{tabular}
\end{center}
\caption[3]{\label{3} As Table I for the motion of the Sun relative to
the LSR, with $\Delta_{{\rm max}}^{{\rm H}}$ the maximum deviation
from $T(v_{{\rm min}})$ found using $v_{\odot, \, {\rm pec}}= (10.0, 5.2,
7.2) \, {\rm km \, s^{-1}}$, as found from Hipparcos data~\cite{hipp}.
The full expression for the Earth's orbit (Ref.\cite{ls}
and eq.~\ref{vels}) is used here.}
\end{table}

\begin{table}
\label{tab4}
\begin{center}
\begin{tabular}{|c|c|c|c|}
$v_{\odot, \, {\rm pec}} ({\rm km \, s^{-1}}$) & $t_{{\rm p}}$ (days) &
$\Delta_{{\rm max}}^{{\rm hyp}}$ & $\Delta_{{\rm max}}^{{\rm
sinsoid}}$ \\ \hline \hline
(0, 0, 0) & 159 -- 153 & 2.4\% -- 13.8\% & 0.1\% -- 2.6\% \\ \hline 
(10, 15, 8) & 152 -- 146 & 2.8\% -- 20.1\% & 0.2\% -- 2.5\%\\ \hline 
(9, 12, 7) & 153 -- 147 & 1.9\% -- 13.6\% & 0.2\% -- 2.8\% \\ \hline
(10.0, 5.2, 7.2) & 152 -- 143 & n/a & 0.3 \% -- 3.2\% \\ 
\end{tabular}
\end{center}
\caption[4]{\label{4} As Table III for the fiducial triaxial halo
model.
}
\end{table}

To provide a quantitative comparison tables I and II contain the day
at which $T(v_{{\rm min}}, \, t)$ (and hence the differential event rate)
is largest, $t_{{\rm p}}$, the maximum deviation from $T(v_{{\rm
min}}, \, t)$ calculated using the full expression for the Earth's motion
(eq.~(\ref{vels})), and the maximum deviation from a sinusoidally
varying $T(v_{{\rm min}}, \, t)$ with the same mean, amplitude and
$t_{{\rm p}}$, for the standard halo model and the fiducial triaxial
model respectively.  Tables III and IV contain the same data for
$T(v_{{\rm min}}, \, t)$ found using each of the sets of values for the
Sun's velocity with respect to the LSR. For reference the DAMA
collaboration have carried out a fit to the modulation of their data
using the function $A \cos{ \left[ 2 \pi (t- t_{{\rm p}})/ T
\right]}$, and when $T$ is fixed at one year they found $t_{{\rm p}}=
144 \pm 13$ days~\cite{dama}.

The time at which the event rate is maximum~\footnote{For simplicity
we neglect the ``flip'' in the phase which occurs at small $v_{{\rm
min}}$ in this discussion.}  varies by up to ten days depending on the
expression used for the Earth's velocity, and depends on $v_{{\rm
min}}$ if the velocity distribution is anisotropic~\cite{ck,fs}.  If
only the component of the Earth's velocity in the Y direction is used,
then the change in the phase which occurs for anisotropic halos is
missed. The three sophisticated expressions produce results which are
in reasonably good agreement however, with maximum deviations of
around a few per-cent for $v_{{\rm min}} \lesssim 400 \, {\rm km \,
s^{-1}}$. For larger $v_{{\rm min}}$ the exact form of the high energy
tail of the speed distribution becomes important, so that using a
different expression for the Earth's velocity produces a large
fractional change in $T(v_{{\rm min}}, \, t)$. As the mean event rate
is small for large $v_{{\rm min}}$, the absolute errors are small
though and therefore not so important from an experimental point of
view.

The maximum deviations from a sinusoidal $T(v_{{\rm min}}, \, t)$ are
no more than a few per-cent for $v_{{\rm min}} \lesssim 400 \, {\rm km
\, s^{-1}}$, as is expected since the error in neglecting the 2nd and
higher order terms in the Taylor expansion is ${\cal O}((v_{{\rm e,
Y}}/v_{\odot})^2 \sim 0.01)$. The amplitude of the annual modulation
is largest for large $v_{{\rm min}}$~\cite{amtheory}, as in the tail of
the speed distribution the fraction of particles with speed greater
than some fixed value changes significantly with $v_{{\rm e}}(t)$, the
Taylor expansion is then inappropriate and the deviation from
sinusoidal variation becomes large (up to around $10\%$). For the
fiducial anisotropic model we have chosen the deviations happen to be
smaller than for the standard isotropic halo model, but this is not
the case in general.

\begin{figure}[t]
\centering \includegraphics[width=0.45\textwidth]{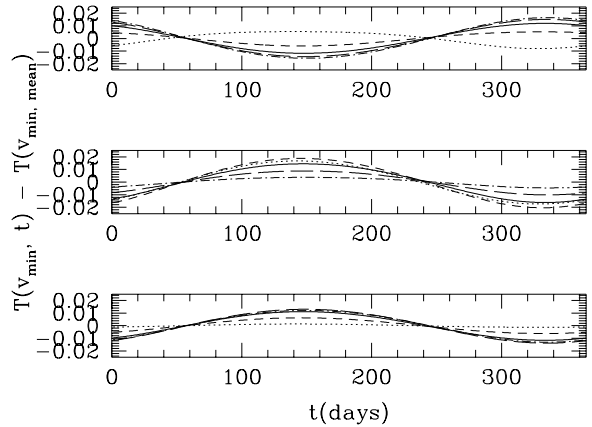}
\caption[fig6]{\label{fig6} The dependence of $T(v_{{\rm min}}, \, t)$ on
the contribution of the halo to the local circular velocity, $v_{{\rm
c, h}}$ for the standard halo model for (from top to bottom) $v_{{\rm
min}}=100, 200$ and $300 \, {\rm \, km s^{-1}}$ and $v_{{\rm c, \, h}}=
100$ (dotted line), $125$ (short dashed), $150$ (solid), $175$ (long
dashed) and $200 \, {\rm km \, s^{-1}} $ (dot-dashed).}
\end{figure}

The time at which the event rate is maximum is relatively weakly
dependent on the Sun's velocity with respect to the LSR (provided that
it isn't neglected entirely), however the maximum deviation from
$T(v_{{\rm min}}, t)$ found using the Hipparcos values for the Sun's
velocity depends strongly on the component in the Y direction and can
be large.  Finally we note that combining a poor approximation for the
Earth's orbit with an erroneous value for the Sun's velocity with
respect to the LSR would produce even larger errors.

\subsection{Local WIMP velocity distribution}

The region of WIMP mass-cross-section parameter space consistent with
the DAMA annual modulation signal depends strongly on the contribution
of the halo to the circular velocity at the solar radius, $v_{{\rm c,
\, h}}(R_{0})$~\cite{amdama}. To explicitly illustrate the reason for
this, in Fig.~6 we plot $T(v_{{\rm min}}, \, t)$ for $v_{{\rm
min}}=100, 200$ and $300 \, {\rm km \, s^{-1}}$ for values of $v_{{\rm
c, \, h}}$ in the range $100-200 \, {\rm km \, s^{-1}}$, for the
standard halo model. As $v_{{\rm c, \, h}}$ (and hence the typical
WIMP velocity) is decreased the value of $v_{{\rm min}}$ at which the
modulation flips phase and also that at which the amplitude of the
modulation reaches a local maximum, are both smaller. For instance for
$v_{{\rm c, \, h}}=100 \, {\rm km \, s^{-1}}$ the local maximum occurs
at roughly $v_{{\rm min}}= 155 \, {\rm km \, s^{-1}} $ compared with
$v_{{\rm min}}= 220 \, {\rm km \, s^{-1}} $ for $v_{{\rm c, \, h}}=150
\, {\rm km \, s^{-1}}$.

\begin{figure}[t]
\centering
\includegraphics[width=0.45\textwidth]{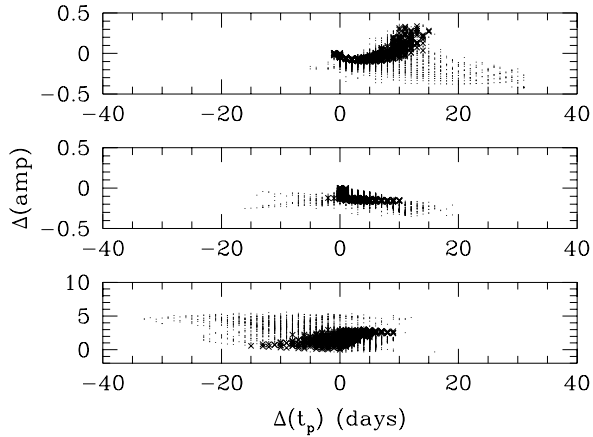}
\caption[fig7]{\label{fig7} The shift in the phase, $\Delta(t_{{\rm
p}})$, and the fractional change in the amplitude of $T(v_{{\rm min}},
\, t)$, $\Delta({\rm amp})$, (relative to the standard isotropic halo
model) for the logarithmic ellipsoidal model with the Sun located on
the intermediate axis, for parameter values for which the ratio of any
two of the velocity dispersions is no greater than 1:3 (dots) and for
which also the anisotropy parameter is in the range $0 < \beta <0.4$
and the axes ratios are $0.6 < I_{1} < 1$ and $0.3 < I_{2} < 1$
(crosses).  From top to bottom $v_{{\rm min}}=100, 300, 500 \, {\rm km
\, s^{-1}}$.  The contribution of the halo to the local circular
velocity is fixed at $v_{{\rm c, \, h}}= 150 \, {\rm km \, s^{-1}} $.}
\end{figure}

\begin{figure}[t]
\centering
\includegraphics[width=0.45\textwidth]{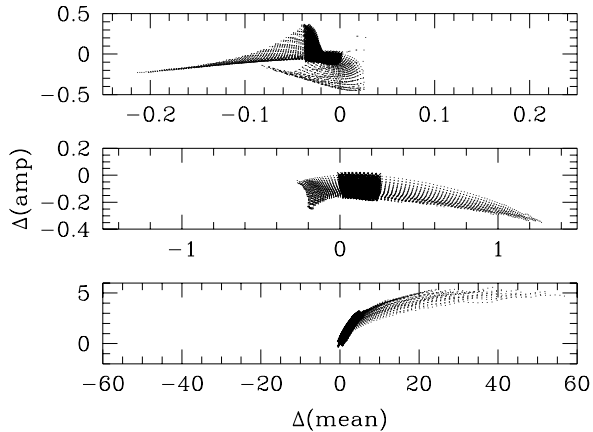}
\caption[fig8]{\label{fig8} As Fig.~7 for the fractional change in the
mean, $\Delta({\rm mean})$, and amplitude, $\Delta({\rm amp})$, of
$T(v_{{\rm min}}, \, t)$ (relative to the standard isotropic halo
model).}
\end{figure}

As our aim is to examine how physical and observational constraints
restrict changes in the phase and shape of the annual modulation
signal (rather than to carry out a detailed comparison with
experimental data) we focus on the logarithmic ellipsoidal model,
which reproduces some of the important features (triaxiality and
velocity anisotropy) of real galaxy halos. We examine the form of
$T(v_{{\rm min}}, \, t)$, with the Sun located on the intermediate
axis, for the sets of values of the velocity dispersions that are
unlikely to lead to instabilities and those which also correspond to
halos with axis ratios and velocity anisotropy consistent with
observed and simulated halos, as discussed in Sec. III.  In Fig.~7 we
plot the shift in the time at which $T(v_{{\rm min}}, \, t)$ is
maximum, $\Delta t_{{\rm p}}$, and the fractional change in the
amplitude of the variation of $T(v_{{\rm min}}, \, t)$ relative to
that for the standard isotropic halo model, for $v_{{\rm min}}=100,
300$ and $ 500 \, {\rm km \, s^{-1}}$, fixing $v_{{\rm c, \, h}}= 150
\, {\rm km \, s^{-1}}$.  In Fig. 8 we plot the fractional changes in
the mean and amplitude of the variation of $T(v_{{\rm min}}, \, t)$,
relative to the standard isotropic halo model.
 
If we consider all sets of velocity dispersions for which
$\sigma_{{\rm j}}/3 < \sigma_{{\rm i}} < 3 \sigma_{{\rm j}}$, the
shift in $t_{{\rm p}}$ has magnitude of up to 40 days and is quite
strongly dependent on $v_{{\rm min}}$. The change in the amplitude is
of order $10$s of per-cent for $v_{{\rm min}}=100$ and $300 \, {\rm km
\, s^{-1}}$ and up to a factor of 5 for $v_{{\rm min}}=500 \, {\rm km
\, s^{-1}}$, while the change in the mean is even larger increasing
from $10$s of per-cent for $v_{{\rm min}}=100 \, {\rm km \, s^{-1}}$
to more than an order of magnitude at $v_{{\rm min}}=500 \, {\rm km \,
s^{-1}}$. The large change in the mean and amplitude for large
$v_{{\rm min}}$ occur because the logarithmic ellipsoidal model has a
more extended tail of high velocity particles than the standard halo
model~\cite{newevans,myhalomod} however, as noted in Sec. III, the
mean event rate is very small for large $v_{{\rm min}}$. When only
sets of velocity dispersions which correspond to halos with realistic
axis ratios and velocity anisotropy are considered the maximum shift
in $t_{{\rm p}}$ is roughly 50 $\%$ smaller and the the maximum change
in the mean signal is roughly an order of magnitude smaller, which
illustrates the importance of only considering velocity distributions
that are physically and/or observationally reasonable. Note that the
standard isotropic halo model (which has $\beta=0$ and
$I_{1}=I_{2}=1$) lies at the edge of the observational constraints we
impose.

In Figs.~9 and 10 we plot the shifts in the mean, amplitude and phase
relative to the standard isotropic halo model with $v_{{\rm c, \, h}}=
150 \, {\rm km \, s^{-1}} $, for the same sets of velocity dispersion
ratios but now with $v_{{\rm c, \, h}}= 200 \, {\rm km \, s^{-1}} $.
The changes in the signal due to the change in $v_{{\rm c, \, h}}$ are
of the same order of magnitude as the changes from varying the other
parameters of the model.

We do not find shifts in $t_{{\rm p}}$ as large as those found by Copi
and Krauss~\cite{ck} and Fornengo and Scopel~\cite{fs}.  For many of
the sets of parameter values we have considered the shift in $t_{{\rm
p}}$ is large enough to be incompatible with the phase of the DAMA
annual modulation signal, however. This suggests that if all three
components of the Earth's velocity are included when calculating the
event rate, then the consideration of anisotropic halo models may not
lead to as large an increase in the region of WIMP mass and
cross-section parameter space corresponding to the DAMA annual
modulation signal as found in Ref.~\cite{damare}. As the DAMA data is
not publicly available, however, it is not possible to confirm this.

\begin{figure}[t]
\centering
\includegraphics[width=0.45\textwidth]{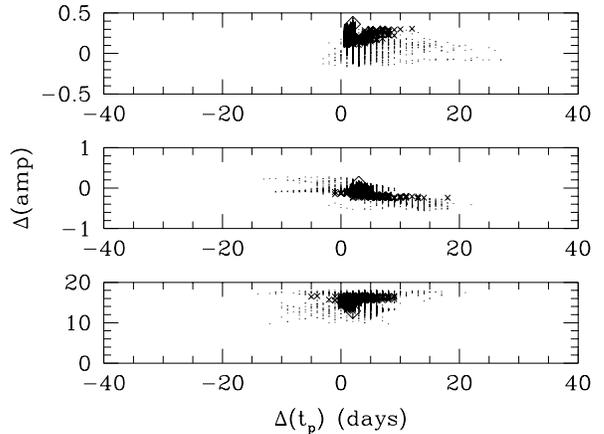}
\caption[fig9]{\label{fig9} As Fig.~7 for the logarithmic ellipsoidal
model with $v_{{\rm c, \, h}}= 200 \, {\rm km \, s^{-1}} $, compared
to the standard isotropic halo with $v_{{\rm c, \, h}}= 150 \, {\rm km
\, s^{-1}} $.  The diamonds denote the standard isotropic halo with
$v_{{\rm c, \, h}}= 200 \, {\rm km \, s^{-1}}$. }
\end{figure}

\begin{figure}[t]
\centering
\includegraphics[width=0.45\textwidth]{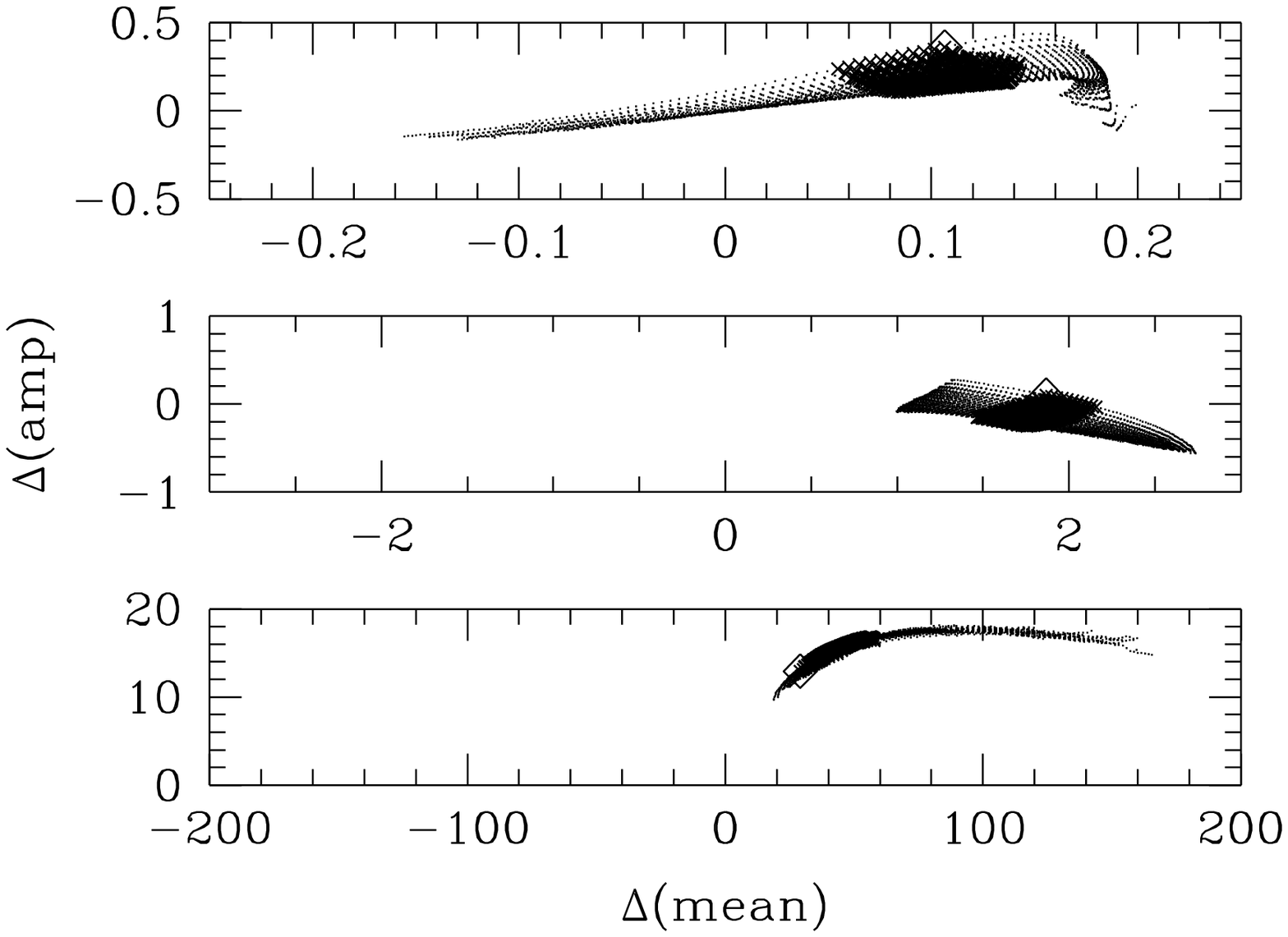}
\caption[fig10]{\label{fig10} As Fig.~8 for the logarithmic
ellipsoidal model with $v_{{\rm c, \, h}}= 200 \, {\rm km \, s^{-1}} $,
compared to the standard isotropic halo with $v_{{\rm c, \, h}}= 150 \,
{\rm km \, s^{-1}} $.  The diamonds denote the standard isotropic halo
with $v_{{\rm c, \, h}}= 200 \, {\rm km \, s^{-1}}$.}
\end{figure}

Finally we overview the effect of streams of particles on the annual
modulation signal. A stream of particles with negligible velocity
dispersion would produce a contribution to $T(v_{{\rm min}}, \, t)$ which is
a step~\footnote{The velocity dispersion would in fact be finite, but
small, and would lead to a broadening in the step~\cite{swf}.}, the
position and amplitude of which varies with time~\cite{swf}:
\begin{equation}
\label{tqclump}
 T_{{\rm s}}(v_{{\rm min}}, \, t) =    \left\{ \begin{array}{ll}
      \displaystyle
         \frac{\sqrt{\pi} \, v_{{\rm c, \, h}}}{2} \,
        \frac{\rho_{{\rm s}}}{\zeta \, \rho_{0.3}} \,
             \frac{1}{v_{{\rm s}}(t)} \,, 
      & {\rm for~} v_{{\rm min}} < v_{{\rm s}}(t) \,,
      \\ 0 ,
      & {\rm for~} v_{{\rm min}} > v_{{\rm s}}(t) \,.
  \end{array} \right.
\end{equation}
where $\rho_{{\rm s}}$ is the density of the stream and $v_{{\rm
s}}(t)$ is the speed of the stream in the rest frame of the detector,
which is time dependent due to the time dependence of the Earth's
velocity. The variation of the amplitude and position of the step
depend on the relative orientation of the Earth's orbit and the orbit
of the clump; for a clump of fixed velocity the closer the alignment
of the clump's path with the Earth's orbit the larger the variation in
$v_{{\rm s}}(t)$ (in the, extremely improbably, case that the clump's
path was perpendicular to the Earth's orbit there would be no
variation).  As outlined in Ref.~\cite{swf} if such a step were
detected in the differential energy spectrum then the density, speed
and direction of the stream responsible could, in principle (with a
large number of events and a detector with good energy resolution), be
recovered. Multiple streams would produce multiple steps at different
energies with different temporal variations, which might be difficult
to disentangle.  Even if most of the dark matter is smoothly
distributed a high velocity ($v_{{\rm s}} \gtrsim 500 \, {\rm km
s^{-1}}$) stream of particles from a late accreting clump could
produce a step in the differential event rate at large energy, the
position of which would be modulated as discussed
above~\cite{swf}. However as the event rate decreases roughly
exponentially with increasing energy this would only be detectable
with very large statistics (large target mass and exposure). A similar
signal would be produced if there is a population of extragalactic
WIMPs~\cite{fgs}.

\section{Discussion}

We have investigated the effect of uncertainties in astrophysical
inputs (the motion of the Earth with respect to the Galactic rest
frame and the local WIMP velocity distribution) on the calculation of
the WIMP annual modulation signal.  Accurate calculation of the shape
and phase of the annual modulation signal requires all three
components of the Earth's velocity with respect to the Sun to be taken
into account properly. If the components perpendicular to the Sun's
motion are neglected, the (energy dependent) shift in $t_{{\rm p}}$
which occurs if the local velocity distribution is anisotropic is
missed. Neglecting the motion of the Sun with respect to the Local
Standard of rest, $v_{\odot, \, {\rm pec}}$, leads to an error of
around 10 days in $t_{{\rm p}}$, and uncertainties in $v_{\odot, \,
{\rm pec}}$ lead to an error of order a few days in $t_{{\rm p}}$ (as
stated in Ref.~\cite{gg}) and errors in the shape of the signal which
grow from a few percent at low energies to more than ten per-cent at
high energy. Approximating the modulation with a sinusoid with the
same mean, amplitude and phase produces errors of up to ten
per-cent. The errors in $t_{{\rm p}}$ are crucial when comparing
theoretical expectations with the modulation observed by the DAMA
collaboration~\cite{dama} (which has phase $t_{{\rm p}}= 144 \pm 13$
days). The errors in the shape of the signal induced by uncertainties
in the Earth's motion are currently less important, but may become
important for annual modulation searches with tonne size detectors
(such as the planned GENIUS detector~\cite{genius}), especially if we
want to extract information about the local velocity distribution from
an observed signal.

If the local WIMP velocity is anisotropic then the phase~\cite{ck,fs},
amplitude~\cite{amgen,amme,gg} and even shape~\cite{gg,fs} of the
annual modulation signal can change. We have investigated, focusing on
the logarithmic ellipsoidal halo model, how the form of the annual
modulation changes for parameter choices which correspond to
physically and observationally reasonable halo models. We found that
for reasonable sets of values for the velocity dispersions the shift
in $t_{{\rm p}}$, which is energy dependent, can be up to 20 days and
the mean and amplitude of the signal change by tens of per-cent, at
experimentally accessible energies. It is possible that other halo
models could produce larger changes in the annual modulation signal,
however we have shown, in the context of this model, that restricting
the choice of parameter values to those that are physically and
observationally reasonable seriously restricts the changes in the
signal. We have also argued that a multivariate Gaussian velocity
distribution, with axes corresponding to the axes of the halo, and
velocity dispersions with large ratios is unlikely to correspond to a
physically reasonable halo model.

It is often assumed that the contribution of the luminous components
of the MW to the circular velocity at the solar radius is negligible
so that $v_{{\rm c, h}}(R_{\odot}) \sim v_{{\rm c}}(R_{\odot}) \approx
220 \, {\rm km s^{-1}}$. This is not the case and in fact there is a
large uncertainty in the value of $v_{{\rm c, \,
h}}(R_{\odot})$~\cite{massmod,mooredm}, which is important since (as
we saw in Sec. IIIB) the mean and amplitude of the event rate (and
hence the region of WIMP mass- cross-section parameters space
corresponding to an observed modulation~\cite{amdama,damare}) depend
quite sensitively on $v_{{\rm c, \, h}}(R_{\odot})$.

Finally we discussed the possibility that the local WIMP distribution
is not completely smooth~\cite{mooredm,swf,swnew}. If the local dark
matter distribution consists of a number of streams of particles with
small velocity dispersion, then the event rate would contain a number
of steps, the amplitude and position of which would vary with
time~\cite{swnew}.  Even if there is a smooth background WIMP
distribution, high velocity streams from late accreting
subhalos~\cite{swf,hws} may be detectable with good energy resolution
and a large number of events.

In summary, it is important to properly take into account
astrophysical uncertainties (not just in the WIMP velocity
distribution, but also in the motion of the detector with respect to
the Galactic rest frame) when calculating the WIMP annual modulation
signal. Analyzing data assuming a sinusoidal modulation with fixed
phase could lead to erroneous constraints on, or best fit values, for
the WIMP mass or cross-section, even worse a WIMP signal could be
overlooked. On the other hand using unrealistic halo models or
parameter values could lead to overly restrictive exclusion limits or
a misleadingly large range of allowed values of the WIMP mass and
cross-section.

\section*{Acknowledgments}

A.M.G.~was supported by the Swedish Research Council.


\appendix
\section{Erratum}

While the halo only contributes some fraction of the circular velocity
at the solar radius, it is the total circular velocity $v_{{\rm
c}}(R_{0}) = 220 \pm 20 {\rm km s^{-1}}$ (and not just the halo's
contribution) which determines the velocity dispersions of the dark
matter particles; the dark matter particles feel the gravity of all
the components of the MW. Some of the values of $v_{{\rm c}}(R_{0})$
used for quantitative calculations in this paper are therefore too
low. The main conclusions of this paper, regarding the effects of
uncertainties in the modelling of the Earths's motion and the local
WIMP velocity distribution on the phase and shape of the annual
modulation signal, are unchanged however.

I am grateful to Piero Ullio and Joakim Edsj\"{o} for bringing this
issue to my attention.

\end{document}